\documentclass{ws-procs9x6}

\begin{document}

\title{Two loop renormalisation of the magnetic coupling in hot QCD and spatial Wilson loop}

\author{P. GIOVANNANGELI}
\address{Centre de Physique Th\'eorique\\
163, avenue de Luminy, \\ 
13288, Marseille cedex 09, France\\ 
E-mail: giovanna@cpt.univ-mrs.fr}

\maketitle

\abstracts{With the known two loop renormalisation of the magnetic coupling, the 4D 
results of the spatial Wilson loop are compared to the prediction from the magnetostatic sector.}

\section{Effective theories for hot QCD}

Perturbation theory in hot QCD suffer from infrared (IR) divergences.
These divergences can be best confronted by constructing
effective field theories for the low-energy dynamics. 
%For general
%real-time observables the relevant theory is called the
%Hard Thermal Loop (HTL) action~\cite{Pisarski:1988vd,Braaten:1989mz},
%while 
For the partition function and other static observables, the relevant theory is
the Dimensionally Reduced (DR) effective
theory~\cite{Ginsparg:1980ef,Appelquist:vg}.
%Both theories encompass the same physics:
%the dynamical (Debye) screening of electrostatic fields in the plasma

%gives a thermal mass to the electrostatic potential and cuts off
%the corresponding IR divergences.
The main idea of DR is to separate systematically the physical
scales of the quark-gluon plasma and to construct, for each of these scales,
an effective action. The procedure is as follows:

\begin{itemize}

\item At distances of order $1 / T$,
      physics is described by the non-static modes. But, if one wants
      to go to larger distances, one meets with IR divergences,
      due to massless static modes.

\item At distances of order $1 / gT$, however,
      electrostatic modes get screened by the Debye mechanism,
      like in a classical electromagnetic plasma. This thermal
      mass is going to regulate the electric IR divergences. At these 
      scale, QCD is described by a 3D electrostatic Lagrangian $L_{E}$ 
      that reads:

\begin{eqnarray}
L_{E} & = &  Tr(\vec D(A)A_0)^2+m_E^2TrA_0^2
                     +\lambda_E(Tr(A_0^2))^2+ {} \nonumber\\
              & + &  \bar\lambda_E \big(Tr(A_0)^4-{1\over 2}(TrA_0^2)^2\big)
                     +{1\over 2}Tr F_{ij}^2+\delta L_E.
\label{estat}
\end{eqnarray}

\noindent with as parameters $g_E( g, T)$, $m_E (g, T)$, ...
      and $\delta L_E$ represents higher orders operators 
      of relative order $g^4$.

\item Furthermore, at distances of order $1 / g^2T$, magnetostatic
      modes are the dominant ones. At these distances QCD
      is described by a confining 3D Yang-Mills theory 
      $L_{M}$ that is non-perturbative.

\begin{eqnarray}
L_{M} & = & {1\over 2}Tr F_{ij}^2+\delta L_M.
\label{mstat}
\end{eqnarray}

\noindent with a gauge coupling $g_M( g_E, m_E,$...). 
$\delta L_M$ is of relative order  $g^3$.

\end{itemize}

If one wants to compute the thermal average of an observable, then
each of these three scales will have its own contribution to the final
result.

%Letting $\mathcal{O}$ be an observable, its thermal average reads
%$\langle \mathcal{O} \rangle_T = \mathcal{O}_H + \mathcal{O}_E +
%\mathcal{O}_M$. Here, $\mathcal{O}_H$ is the contribution due to hard modes,
%$\mathcal{O}_E$ the contribution due to electrostatic modes
%(with momenta of order $g T$), while $\mathcal{O}_M$ is the
%contribution of magnetostatic modes (with momenta of order $g^2 T$).
%The parts $\mathcal{O}_E$ and $\mathcal{O}_M$ are
%computed with the 3D Lagrangians $\mathcal{L}_E$ and $\mathcal{L}_M$,
%respectively. 

From the relative order of the truncation in (\ref{estat}) and (\ref{mstat}) we see that $g_M$ is needed to relative order $g^2$. In the next section, 
I will briefly present the result for  the coupling $g_M$
of the 3D magnetostatic Yang-Mills theory $L_M$ at two loop order.

\section{Two-loop determination of the magnetostatic action}

The basic idea behind the effective actions eqns (\ref{estat}) and (\ref{mstat}) 
is that one can compute with both in the region of momenta $p\sim g^2 T$. 
To know what the parameters of the latter are in terms of those of the former
requires computing two-point functions, three point functions etc.  
in both theories and match them. In the matching the diagrams 
of the pure 3d Yang-Mills theory drop out.

Here we will follow a well-known shortcut~\cite{abbott} by introducing a background field
$B_i$ in  $L_E$:
\begin{eqnarray}
 \vec A & = &{1\over {g_E}}\vec B +  \vec Q\nonumber\\
    A_0 & = & g_EQ_0.
\end{eqnarray}
We calculate the fluctuations around the background in a path integral:
\begin{equation}
\exp{-{1\over{g_M^2}}S_M(B)}=\int DQ_0DQ_i\exp{\big(-S_E-{1\over{\xi}}Tr(D_iQ_i)^2\big)}.
\label{path}
\end{equation}

We added a general background gauge term. The resulting action $S_M(B)$ is
gauge invariant to all loop orders and the renormalization of the coupling
is identified from the background field two point function at a momentum $p=O(g^2T)$.

\begin{equation}
 \exp{\big(-{1\over{g_M^2}}S_M(B)\big)}= \exp{\big(-{1\over{g_E^2}}S_M(B)\big)}\big(1+(F_1^{tr}+F_2^{tr}+...)S_M(B)\big).
\label{ident}
\end{equation}

$F_i$ is the sum of all Feynman diagrams  for the two point function of the background field with $i$ loops and reads:

\begin{equation}
F_i=F_i^{tr}~(\delta_{lm}p^2-p_lp_m).
\end{equation}

This leaves us with the relation, using eq.(\ref{ident}):
\begin{equation}
{1\over{g_M^2}}={1\over{g_E^2}}-F_1^{tr}-F_2^{tr},
\end{equation}

with

\begin{eqnarray}
g_E^2F_1^{tr} & = & -{1\over{48}}{g_E^2N\over{\pi m_E}}\\
g_E^2F_2^{tr} & = & -{19\over{4608}}({g_E^2N\over{\pi m_E}})^2.
\label{final}
\end{eqnarray}

This is the main result~\cite{moi2}.

\section{Spatial Wilson Loop}
We want now to test the applicability of 3D physics at medium high T.
Let us consider an observable which has, unlike the pressure whose dominant contribution is the 
Stefan-Boltzmann term due to hard modes, its dominant contribution from the 3D Yang-Mills sector.  
Such an observable can be the spatial Wilson loop in the fundamental representation:
\begin{equation}
W(L)=Tr{P}\exp{(i\oint_L g\vec A \cdot d\vec l)}.
\label{wilsonloop}
\end{equation}

\noindent As $L$ is purely spatial, it measures the magnetic flux in the plasma. The case where  $\vec A$ is in an irreducible representation made of $k$ quarks is described in this volume by C.P. Korthals Altes. The thermal average of this spatial Wilson loop shows area behaviour with a surface tension $\sigma(T)$. As it is a purely magnetic quantity, we expect from dimensional arguments that $\sqrt{\sigma} = c g_M^2$ where $c$ is a nonperturbative proportionality constant.
Indeed, as the average of the loop is due to long distance correlation, hard modes
will not have any effects on the thermal average. In the same way,
for soft modes, we can integrate out the $A_0$ field,
which is what we have done in the previous section while constructing $L_M$,
as the loop does not depend on $A_0$ ! Finally, we have:

\begin{equation}
\langle W(L) \rangle = \exp{-\sigma A(L)}=\frac{\int  D\vec A W(L)\exp{-S_M(A)}}{\int D\vec A\exp{-S_M(A)}},
\label{eq:wloopmagnetic}
\end{equation}

\noindent and it gives, as $\delta L_M$ is of relative order $g^3$,
\begin{equation}
\sigma(T)=c^2g_M^4(1+O(g^3)).
\label{eq:wtensionmagnetic}
\end{equation}

Now the aim is to fit with our formula for $g_M$ the proportionality constant $c$, and to study its extension for finite $T$. For that, we have to go to the lattice. We are going to fit
$ T\over \sqrt{\sigma}$ has a function of $T \over T_c $. Indeed, $g_M$ is a function of $g_E$ and $m_E$ which
are functions of $g$ and $T$.
So, for $N = 3$:

\begin{equation}
g_M^{-2} = g^{-2} T^ {-1} (1 + {g\over 16 \pi} + {19 g^2 \over 512 \pi^2})
\end{equation}

\noindent with

\begin{equation}
g^{-2} = {11 \over 8 \pi^2}(Log( {T \over T_c}) +  Log( {T_c \over \Lambda_{\bar{MS}}}) + 1.90835... ).
\end{equation}

%The $\mu$ dependance will be discussed in next section

So thanks to these formulas, one can fit $ T\over \sqrt{\sigma}$  versus $T \over T_c$
in order to determin  $c$ and $T_c \over \Lambda_{\overline{MS}}$.

\begin{figure}
\begin{center}
\includegraphics{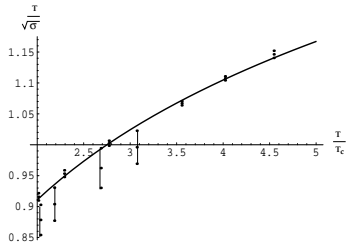}
\caption{Plot of  $T\over \sqrt{\sigma}$  versus $T \over T_c$ and fit with our two-loop formula. Each of the 10 points is shown with its error bar.}
\label{fig:fit1}
\end{center}
\end{figure}

We have taken data points for the $SU(3)$ spatial string tension from
~\cite{Boyd:1996bx,Lutgemeier:xi} for $T > 2 T_c$. So we have $10$ points. The fit is shown in fig~\ref{fig:fit1}.

We have found the following results:

\begin{eqnarray}
{T_c\over{\Lambda_{\overline{MS}}}} &=& 1.78(12)\\
c &=& 0.505(11).
\end{eqnarray}

%\noindent In order to estimate the error in our fit, we have plotted for those $10$ points the confidence levels taken from the $\chi^2$ distribution in
%fig~\ref{fig:quifit}.
%Compared with the results given in the previous chapter, eq.(2.4), our values are incompatible with them.

%\begin{figure}
%\begin{center}
%\includegraphics{clip01.eps}

%\caption{ Plot of various confidence levels in the plane c- $T_c \over \Lambda_{\bar MS}$.
%The inner ellipse delimits the $68$ percent confidence level
%and the outer ellipse the $95$ percent confidence level for the parameters.
%The central values are $c = 0.51$ and  ${T_c \over \Lambda_{\bar{MS}}} = 1.78$.
%}
%\label{fig:quifit}
%\end{center}
%\end{figure}

Then we have also fitted data with $T > 2.5 T_c$. This leaves us with $6$ points.
We obtained
\begin{eqnarray}
{T_c\over{\Lambda_{\overline{MS}}}} &=& 1.57(20)\\
c  &=& 0.488(18)
\end{eqnarray}

%We have plotted for those $6$ points the $95$ percent confidence level contour taken from the $\%chi^2$ distribution.

%\begin{figure}
%\begin{center}
%\includegraphics{sixp.eps}
%\caption{ Plot of the  $95$ percent confidence level (outer ellipse) in the plane c- $T_c \over %\Lambda_{\bar MS}$ for $6$ points
%The central values are $c = 0.488$ and  ${T_c \over \Lambda_{\bar{MS}}} = 1.57$.
%}
%\label{fig:quifit2}
%\end{center}
%\end{figure}

Let us draw our conclusions.

Despite the fact that the region of confidence is larger, the value of 
%${T_c\over{\Lambda_{\overline{MS}}}}$ and 
c is still incompatible with refs.~\cite{Teper:1998te,Lucini:2002wg,Karsch:1994af}.
To see compatibility one obviously needs string data for higher $T$. It might also be recommendable to use other Monte Carlo updates, like the one described in ref.~\cite{HMeyer}, which is a version of the L\"uscher-Weisz algorithm~\cite{LW}.

%%%%%%%%%%%%%%%%%%%%%%%%%%%%%%%%%%%%%%%%%%%%%%%%%%%%%%%%%%%%%%%%%%%%%%%
% 
%Use this if your figures are put in a subdirectory having the same
%name as the main latex file, ie: 
%
%      ws-procs9x6/procs-fig1.eps      
%      ws-procs9x6/procs-fig2.eps      
%      ws-procs9x6/procs-fig3.eps      
%      ws-procs9x6/procs-fig4.eps      
%      etc.
%
%\begin{figure}[htbp] %ORIGINAL SIZE: width=1.4TRUEIN; height=1.5TRUEIN
%\figurebox{}{}{procf1} %100 percent
%\caption{Labeled tree {\it T}.}
%\end{figure}
%
%%%%%%%%%%%%%%%%%%%%%%%%%%%%%%%%%%%%%%%%%%%%%%%%%%%%%%%%%%%%%%%%%%%%%%%

\end{document}